\documentclass[11pt]{article}
\usepackage[utf8]{inputenc}
\usepackage{times}

\usepackage{xcolor} 
\usepackage{hyperref} 
\hypersetup{
  colorlinks=true,
  citecolor=violet,
  linkcolor=red,
  urlcolor=blue
}

\usepackage{balance} 

\usepackage[sort,numbers]{natbib}

\usepackage{amsmath} 
\usepackage{amsthm} 

\usepackage{tabularx} 
\usepackage{graphicx} 
\usepackage{tikz} 
\usetikzlibrary{automata, positioning} 

\usepackage[letterpaper,top=1in, bottom=1in, left=1.5in, right=1.5in]{geometry}

\usepackage{setspace}
\usepackage{enumitem}
\setlist[enumerate,itemize,description]{itemsep=0.5pt,topsep=0.5pt,partopsep=0.5pt}

\usepackage[textwidth=1.5cm]{todonotes}
\definecolor{kentuckyblue}{RGB}{0, 93, 170}

\title{Analyzing Performance Bottlenecks in Zero-Knowledge Proof Based Rollups on Ethereum}
\author{Md. Ahsan Habib \\Tulane University, mhabib@tulane.edu}
\date{December 2024}

\begin{document}

\maketitle

\begin{abstract}
Blockchain technology is rapidly evolving, with scalability remaining one of its most significant challenges. While various solutions have been proposed and continue to be developed, it is essential to consider the blockchain trilemma — balancing scalability, security, and decentralization — when designing new approaches. One promising solution is the zero-knowledge proof (ZKP)-based rollup, implemented on top of Ethereum. However, the performance of these systems is often limited by the efficiency of the ZKP mechanism. This paper explores the performance of ZKP-based rollups, focusing on a solution built using the Hardhat Ethereum development environment. Through detailed analysis, the paper identifies and examines key bottlenecks within the ZKP system, providing insight into potential areas for optimization to enhance scalability and overall system performance.
\end{abstract}

\section{Introduction}
Blockchain is a distributed, decentralized technology that provides a secure, tamper-proof method for recording data across a peer-to-peer (P2P) network. It stores information in a linear sequence of blocks, each containing a timestamp, transaction data, and a cryptographic hash pointer linking it to the previous block, forming an immutable chain. This structure ensures the integrity of the data by making it nearly impossible to alter any information once it is added to the blockchain, as any tampering would require changing the entire sequence of subsequent blocks. It is widely used in various sectors, including finance, healthcare, supply chain management, and public administration, due to its ability to provide transparent, auditable, and secure records without the need for a central authority \cite{zheng2018blockchain}.

Despite its broad range of applications, blockchain faces significant challenges related to scalability due to its append-only structure. As the number of transactions grows, issues such as slower processing times, higher storage requirements, and increasing transaction fees arise. For example, Bitcoin can only handle around 7 transactions per second (TPS), while traditional payment systems like Visa process thousands of TPS. To address this, various solutions have been proposed. However, due to the blockchain trilemma — the need to balance scalability, security, and decentralization — finding an optimal solution is complex. Solutions such as Layer 2 (L2) protocols based on Ethereum (e.g., ZKRollups, Optimistic Rollups), sharding, and improved consensus mechanisms are being developed to increase scalability while maintaining security and decentralization \cite{thibault2022blockchain}.

In this paper, we will explore a rollup protocol based on ZKP \cite{quisquater1989explain}, implemented on the Ethereum blockchain. Our focus will be on measuring the performance of the solution and identifying its bottlenecks. More specifically we will focus on analyzing the ZKP generation and verfication part which is the crucial part.

\section{Related Works}

The study \cite{chaliasos2024analyzing} analyzes the two different real time ZKRollup solutions namely Polygon ZKEVM \cite{polygonzkEVM} and ZKSyn Era \cite{zksyncera} and thoroughly evaluate them. They show that the Polygon ZKEVM solution that employs STARK and FFLONK proofing systems to achieve predictable proving times of approximately 190–200 seconds per batch. It is also fully compatible with the existing EVM. However, it processes smaller batches compared to other solutions. The data availability costs can reach 50\% of the total transaction cost in certain configurations. Despite this trade-off, it focuses on seamless EVM integration and predictable performance. 

By employing Boojum and PLONK-KZG proofing systems, the other zkSync Era supports batch sizes of up to 5,000 transactions, lowering the cost as little as \$0.01 per transaction. Unlike Polygon ZKEVM, zkSync Era reduces data availability costs through state diffs and blob-based data posting enabled by Ethereum’s EIP-4844. Moreover, it exhibits elastic proving times, such as 240 seconds for a batch size of 5,000 transactions which makes it less predictable. However, it focuses on cost reduction and high-throughput. The following Fig. \ref{fig:zkEVMEra} shows the cost breakdown of the Polygon ZKEVM solution and zkSyn Era solution using the real-time data \cite{chaliasos2024analyzing}.

\begin{figure}[htbp]
\centering
\includegraphics[scale=0.55]{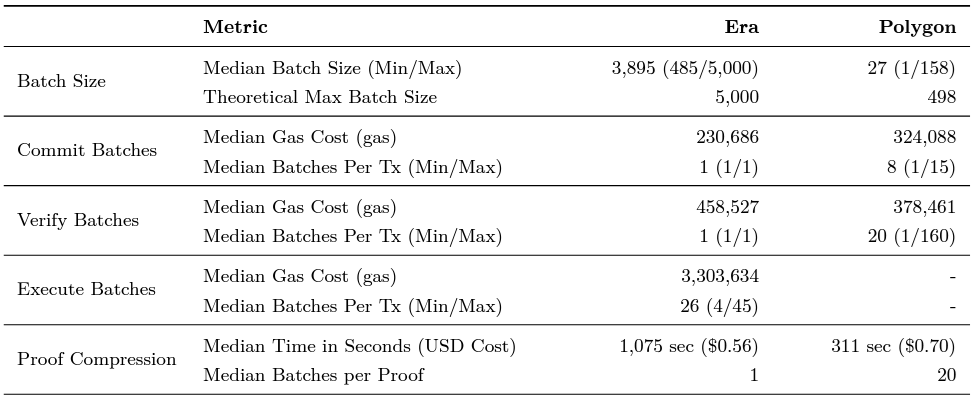}
\caption{Performance of zkSync Era and Polygon ZKEVM using real transaction \cite{chaliasos2024analyzing}.}
\label{fig:zkEVMEra}
\end{figure}

\section{Zero-Knowledge based Rollup}
\begin{figure}[htbp]
\centering
\includegraphics[width=\textwidth]{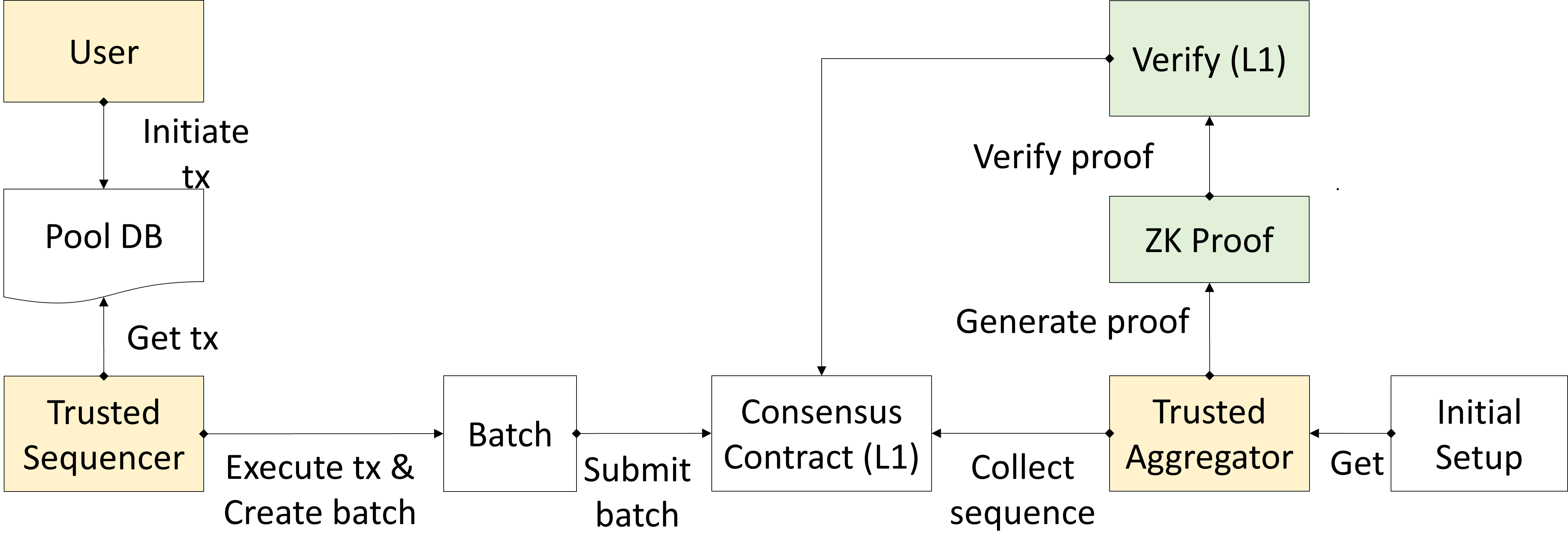}
\caption{Flowchart of a ZKRollup solution.}
\label{fig:zkRollup}
\end{figure}

ZKRollups are a layer-2 (L2) scaling solution for blockchains that leverage ZKP protocols to bundle multiple transactions into a single batch and generate some proof, which is then submitted to the main blockchain. This approach significantly improves scalability by reducing the computational burden on the mainchain while maintaining the security and decentralization of the underlying blockchain. In ZKRollups, transaction data is processed off-chain (L2), and only the proof of validity is posted on-chain (L1), allowing for a high throughput of transactions without overloading the main network. By using cryptographic zero knowledge proof systems, ZKRollups ensure that the transactions are valid. The flow of this process is illustrated in Fig. \ref{fig:zkRollup}, where users initiate transactions and submit them to a database pool (Pool DB). A trusted sequencer then creates a batch by collecting these transactions and sending them to the smart contract. A trusted aggregator subsequently gathers the sequence of batches and generates a proof for each one. Finally, the proof is submitted to the main chain for verification, and if successful, the transactions are validated.

\section{Methodology}
\begin{figure}[htbp]
\centering
\includegraphics[scale=0.5]{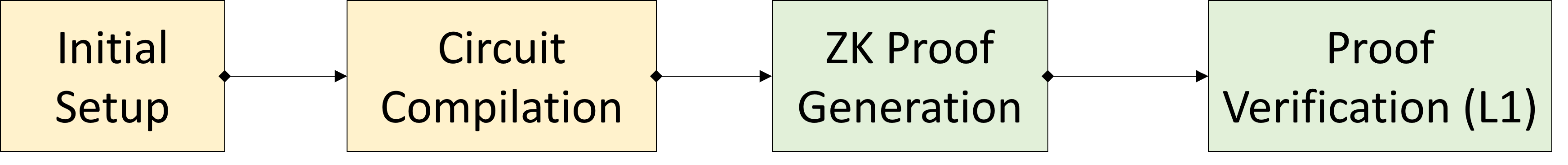}
\caption{Working process of the paper.}
\label{fig:method}
\end{figure}

The focus of this paper is on the critical ZKP part. Specifically, we will analyze the proof generation and proof verification processes for batch creation with varying batch sizes, as well as for the withdrawal circuit, using the Groth16 protocol \cite{groth16zkblog}. Additionally, we will examine the initial trusted setup of the protocol by adjusting different system parameters. To achieve this, we will work through the following four steps, as illustrated in Fig. \ref{fig:method}.

\begin{itemize}
    \item \textbf{Step 1: Initial Setup}: The setup phase includes generating the cryptographic keys and the common reference string (CRS) through a trusted setup ceremony.
    \item \textbf{Step 2: Circuit Compilation}: The Circom \cite{circomdocs} circuit is compiled, transforming the computation into a Rank-1 Constraint System (R1CS) for proof generation.
    \item \textbf{Step 3: ZK Proof Generation}: A ZK Proof is generated off-chain to validate the batch of transactions.
    \item \textbf{Step 4: Proof Verification (L1)}: The proof is submitted to the mainchain (L1), where it is verified by the smart contract to ensure the validity of the off-chain transactions.
\end{itemize}

For more details of these steps, please refer to the section \ref{sec:implementation}.

\section{Implementation}\label{sec:implementation}
The Hardhat Ethereum development environment is a popular framework for building and testing decentralized applications on the Ethereum blockchain. It provides a flexible and efficient environment for developers when implementing ZKP systems like Groth16 within Hardhat. Groth16 is an efficient ZKP protocol known for its small proof sizes and fast verification times, making it ideal for ZKP-based rollups. The libraries like \texttt{snarkjs}, \texttt{circom}, etc. allows seamless interaction with ZKP on the Ethereum network. One can find the similar implementation of this paper here \cite{zkrollupgithub}.

\subsection{Experimental Setup}
To conduct all the computational experiments and benchmarks, we use a personal computer with the following specifications: 

\begin{itemize}
    \item \textbf{Processor}: Intel(R) Core(TM) i7-14700F @ 2.10 GHz
    \item \textbf{RAM}: 32.0 GB
    \item \textbf{System Type}: 64-bit operating system, x64-based processor
\end{itemize}

\subsection{Cost and Performance Analysis}

\begin{figure}[htbp]
\centering
\includegraphics[scale=0.6]{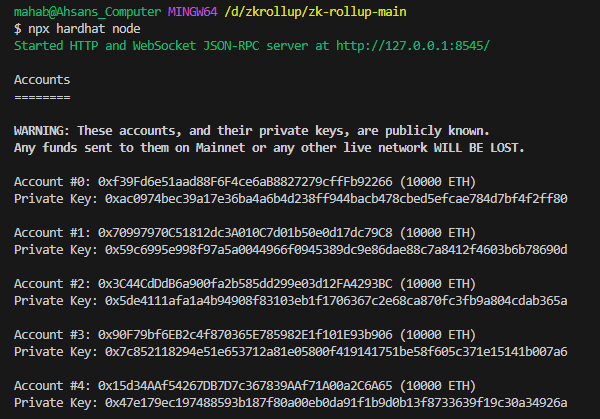}
\caption{Snippet of accounts in Hardhat.}
\label{fig:AccountsHardhat}
\end{figure}

\begin{figure}[htbp]
\centering
\includegraphics[scale=0.45]{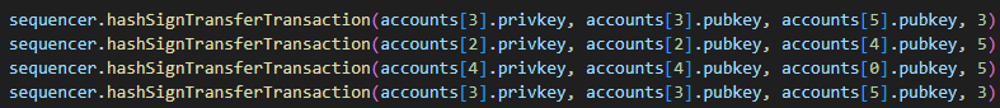}
\caption{Snippet of transactions in Hardhat.}
\label{fig:TxsHardhat}
\end{figure}

In implementing the ZKRollup solution, we evaluate both one-time and repetitive costs associated with the system. The analysis includes key steps in the ZKRollup pipeline, highlighting how these costs scale with varying parameters. To illustrate this, we present some associated Hardhat accounts in Fig. \ref{fig:AccountsHardhat} and some transactions for the creating the batch in Fig. \ref{fig:TxsHardhat}.

\paragraph{One time costs:} The following types of costs are associated with the setup process. Once completed, they do not need to be repeated for subsequent runs.
\begin{itemize}
    \item \textbf{Power of Tau:} The Power of Tau is a multi-party computation process designed to produce a common reference string (CRS) that is required for ZK-SNARKs (Zero-Knowledge Succinct Non-Interactive Arguments of Knowledge). The CRS is used during proof generation and verification. The goal is to ensure that no single participant in the process can compromise the security of the system. It generates a sequence of cryptographic powers of a secret value \( \tau \), such as \( \tau, \tau^2, \tau^3, \dots, \tau^n, \) without revealing \( \tau \). The CRS includes these powers and is used in ZK-SNARKs for encoding the constraints of the system. If we want higher level of security, we have to increase the value of \(n\), hence the cost of this ceremony gets increased, as higher security parameters demand more computation. As shown in Fig. \ref{fig:PowerOfTau}, when the time requirement is exponential with the power of \( \tau \), i.e., in case of \(n\) = 19 it requires more than 30 minutes whereas it requires only 3-4 minutes when we set \(n\) = 15. However, when we set \(n\) = 20, our local machine crashed due to memory run out.

    \begin{figure}[htbp]
    \centering
    \includegraphics[scale=0.5]{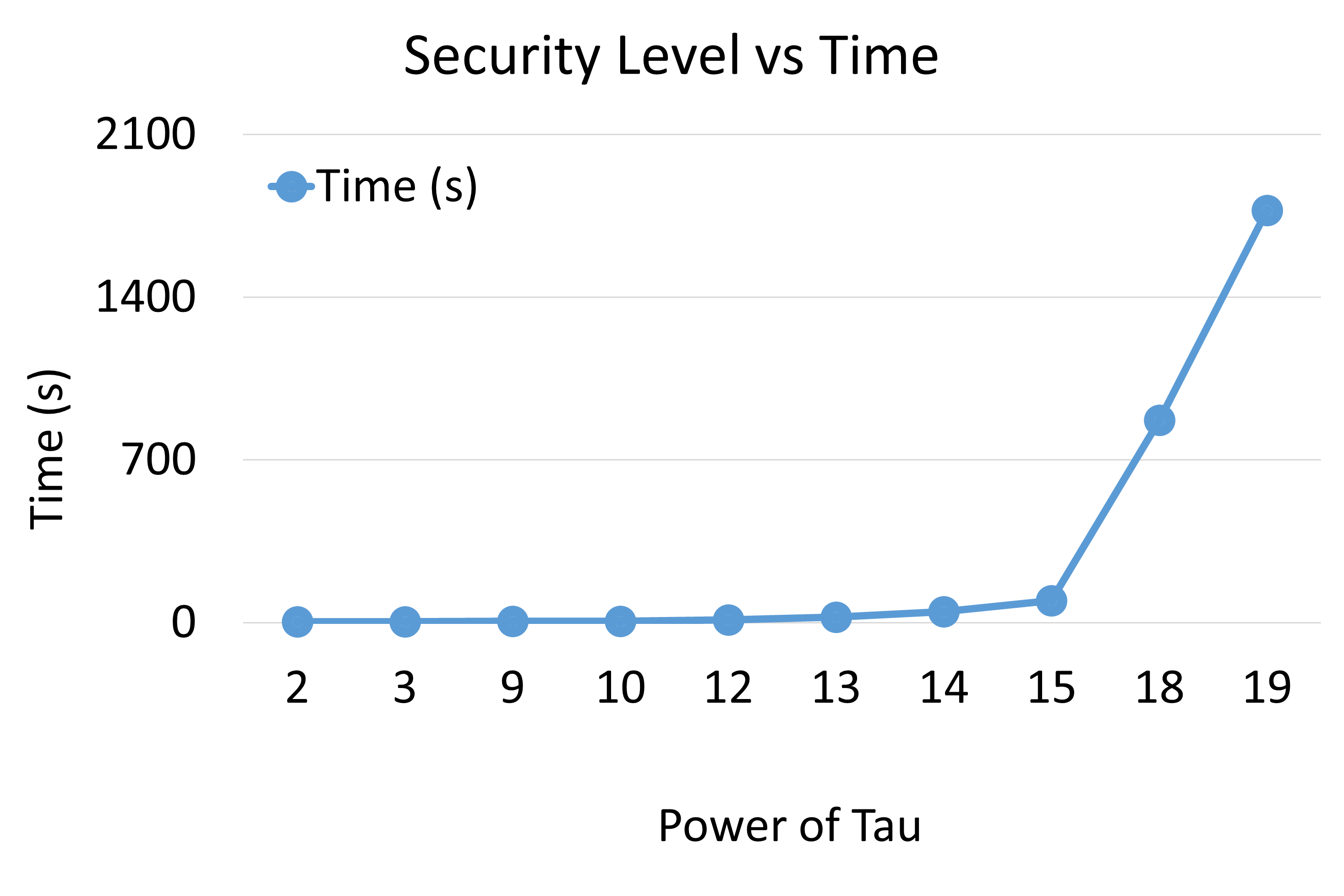}
    \caption{Time requirement for different security level.}
    \label{fig:PowerOfTau}
    \end{figure}

    \item \textbf{Compilation:} The compilation cost of the Circom \cite{circomdocs} circuit for the batch creating increases with larger batch sizes because the complexity of the circuit grows as more transactions are included. In case of batching, the circuit needs to generate and optimize constraints for each transaction, and when the batch size increases, the number of constraints also rises. This results in a greater computational load during the compilation process, as the system must handle and process more data. On the other hand, the withdraw time remains constant because the proof generation process involves the same cryptographic operations. The circuit's overall structure does not change, and hence the proof generation time does not scale directly with batch size. As shown in Fig. \ref{fig:CompileTime}, we test 3 batch sizes with 4, 8 and 16 transactions.

   \begin{figure}[htbp]
    \centering
    \includegraphics[scale=0.5]{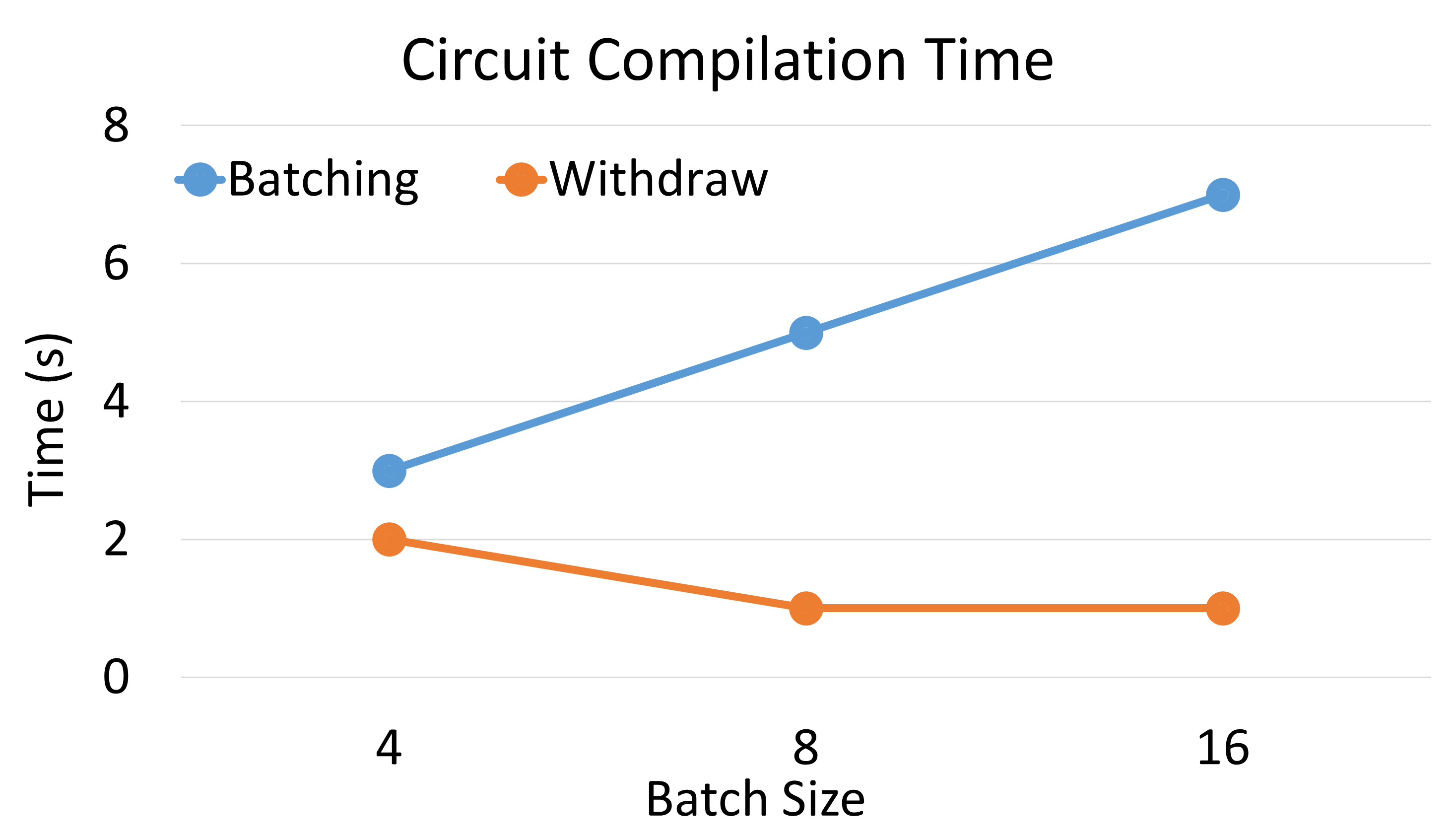}
    \caption{Time requirement for circom circuit compilation for three different batch sizes.}
    \label{fig:CompileTime}
    \end{figure} 
    
    \item \textbf{Key Generation (SNARK):} Similar to compilation, the key generation cost for the circuit for batching the transactions increases with larger batch sizes because generating cryptographic keys for a larger batch involves more computation. As the batch size increases, it directly impacts the complexity of the proving and verification key generation. These keys require complex cryptographic operations such as elliptic curve (EC) pairings and hashing. Consequently, larger batch sizes require longer key generation times. As shown in Fig. \ref{fig:SNARKTime}, here we also test the circuit with three batch sizes.

    \begin{figure}[htbp]
    \centering
    \includegraphics[scale=0.5]{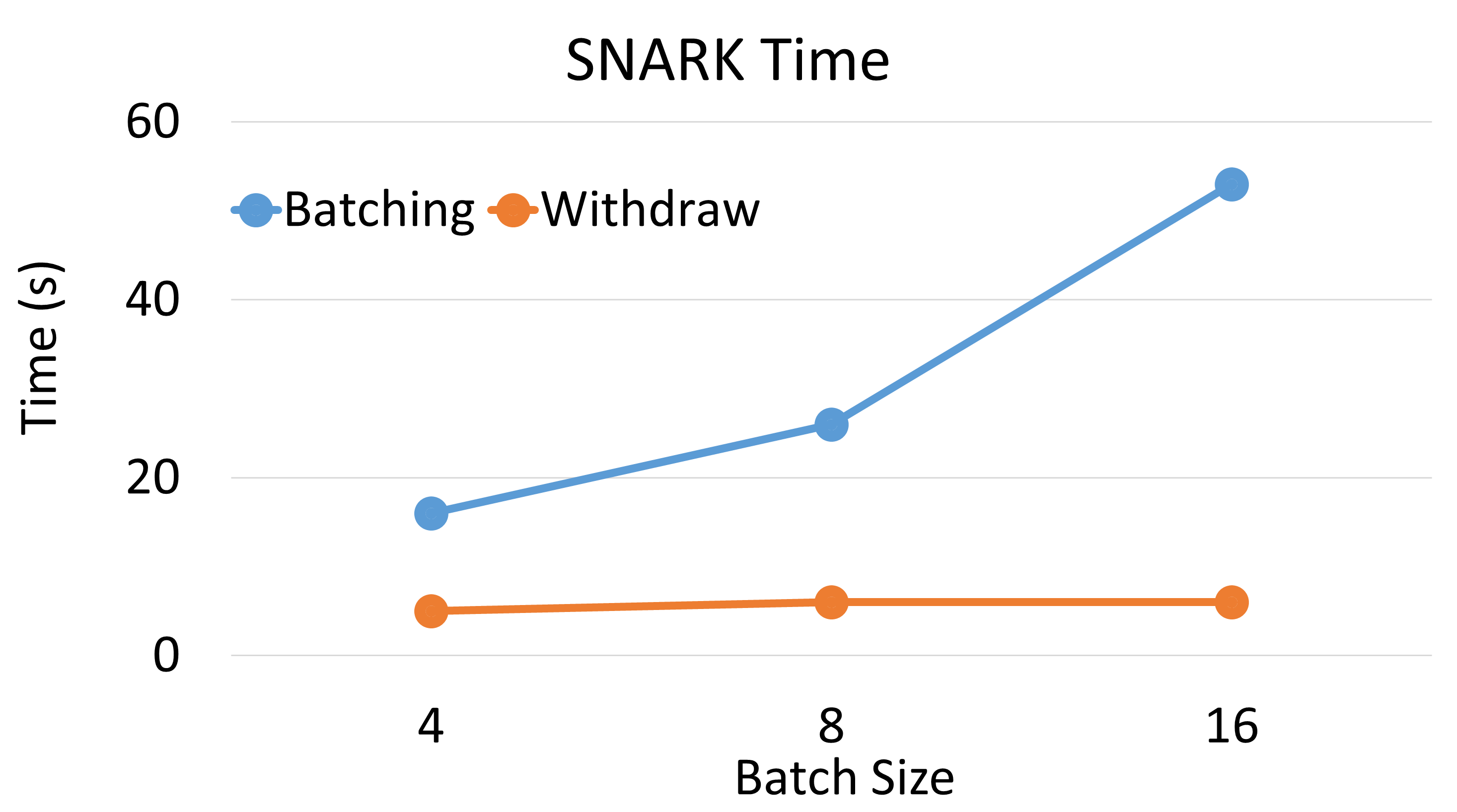}
    \caption{Time requirement for SNARK setup for three different batch sizes.}
    \label{fig:SNARKTime}
    \end{figure}
    
\end{itemize}

\paragraph{Repetitive costs:} Each time we create a batch, we must generate the proof for the batch, which is then verified on the mainchain (L1).
\begin{itemize}
    \item \textbf{Proof Generation:} The cost of generating proofs of batches increases with larger sizes due to processing more transactions. As shown in Fig. \ref{fig:ProofVerifyTime}, the time required for batch proof generation (PrfGenB) increases with higher batch sizes, while the proof generation time for the withdrawal transaction (PrfGenW) remains constant, as it handles only a single transaction regardless of the batch size.
    \item \textbf{Proof Verification:} Verification costs remain consistent, independent of the batch size, benefiting from the succinctness of ZK-SNARK proofs shown in Fig. \ref{fig:ProofVerifyTime}.
    
    \begin{figure}[htbp]
    \centering
    \includegraphics[scale=0.5]{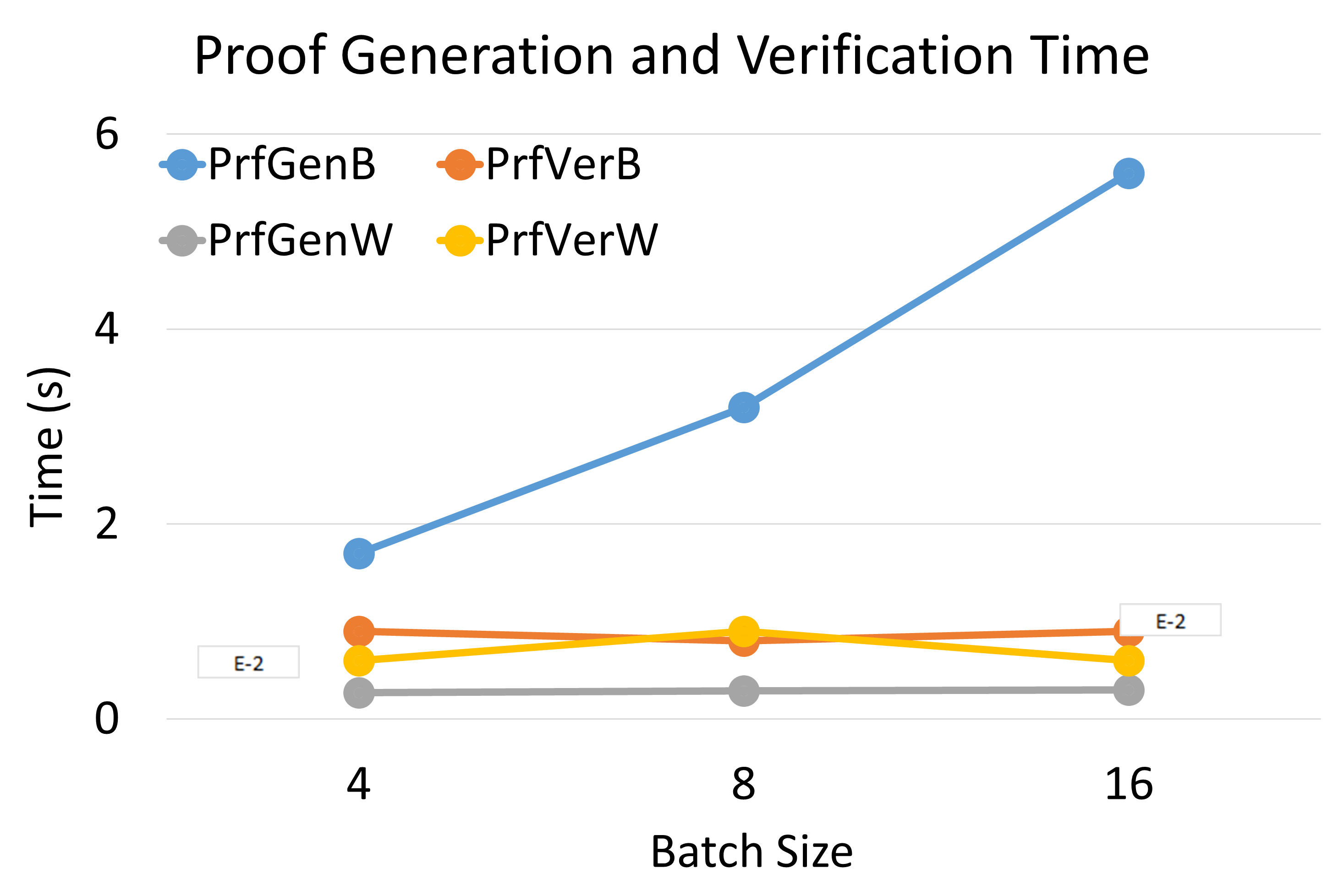}
    \caption{Time requirement for proof generation and verification for different batch sizes.}
    \label{fig:ProofVerifyTime}
    \end{figure}

    \item \textbf{Cost per Transaction:} The cost per transaction decreases with larger batch sizes, reflecting the economies of scale inherent to ZK-Rollup solutions. This is logical, as the overhead of handling the entire batch is shared across all transactions, the cost is lowered. As shown in Fig. \ref{fig:GasPerTx} and Fig. \ref{fig:USDPerTx}, the cost per transaction is illustrated using two metrics: gas used and USD.

    \begin{figure}[htbp]
    \centering
    \includegraphics[scale=0.5]{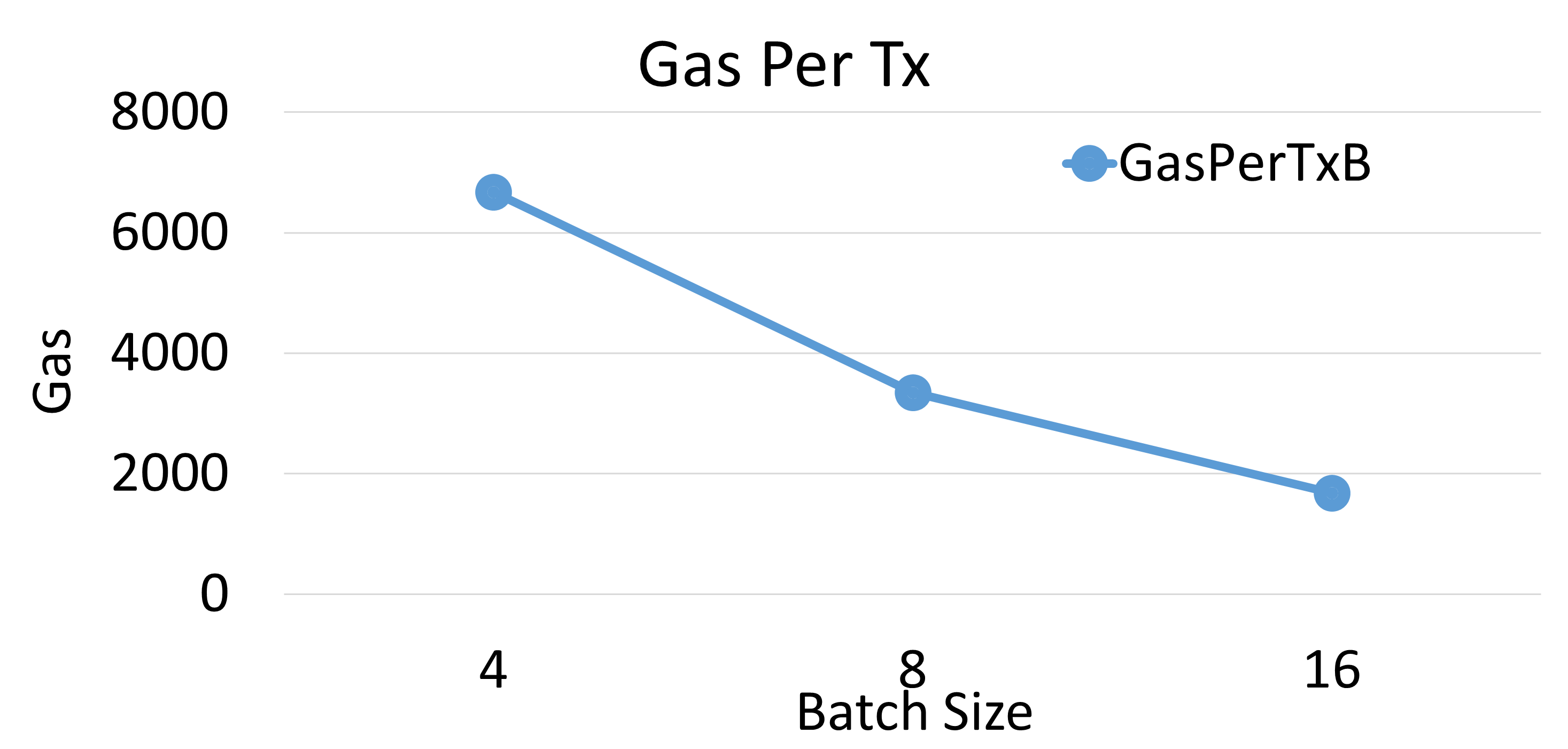}
    \caption{Gas used per transaction for different batch sizes.}
    \label{fig:GasPerTx}
    \end{figure}
    
    \begin{figure}[htbp]
    \centering
    \includegraphics[scale=0.5]{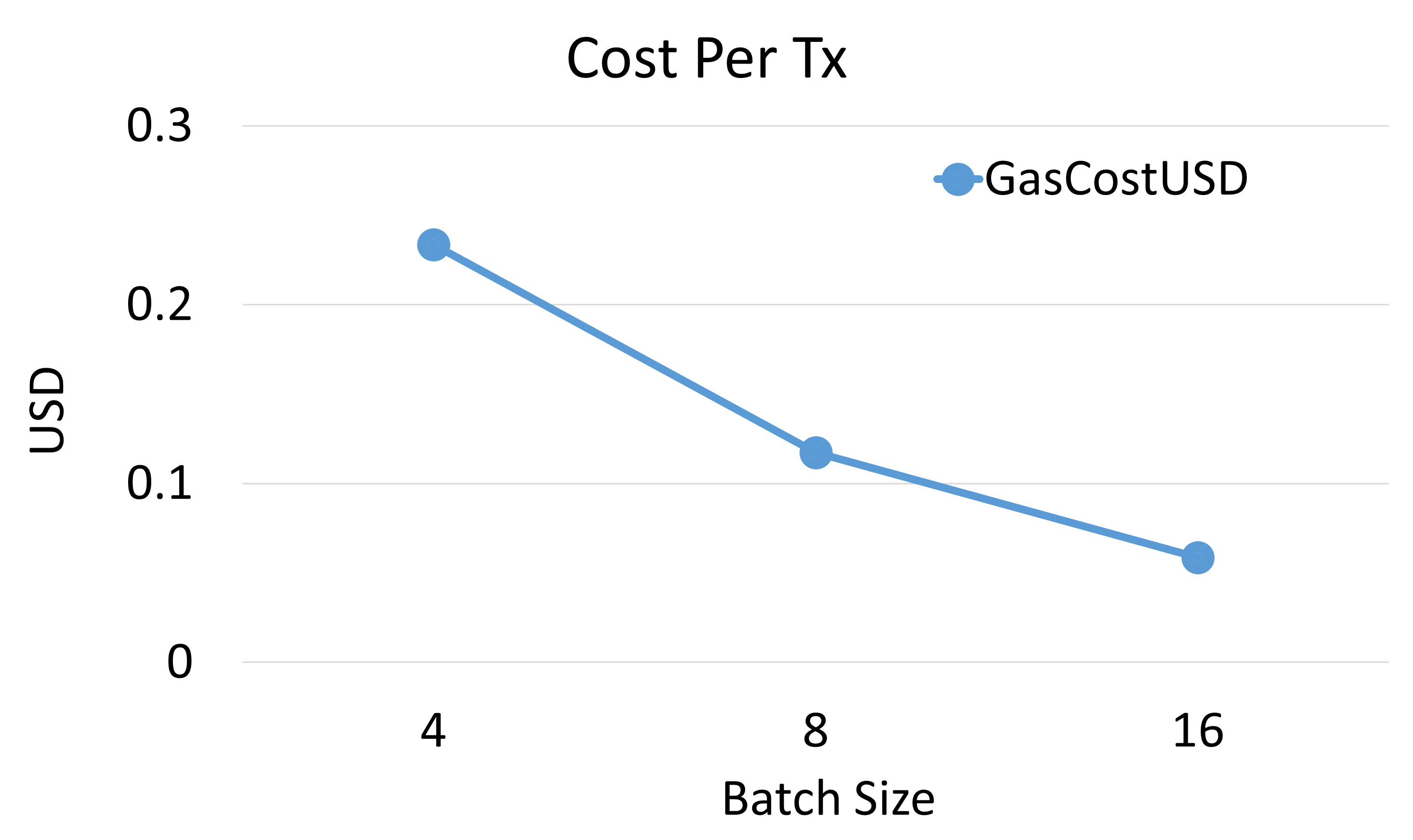}
    \caption{USD cost per transaction for different batch sizes.}
    \label{fig:USDPerTx}
    \end{figure}
    
\end{itemize}

\section{Findings and Future Directions}
From our analysis, we observe that one-time costs, such as Power of Tau, Circom circuit compilation, and SNARK key generation, significantly increase with larger batch sizes due to the growing complexity and resource requirements. However, these costs are incurred only once, and their impact on the overall rollup solution is limited. In contrast, proof verification time remains constant regardless of batch size while the proof generation time increases with larger batches as more transactions add computational complexity.

The primary bottleneck in the zkRollup solution is the increasing proof generation time associated with larger batch sizes. Although economies of scale reduce the per-transaction cost, the computational complexity of generating proofs for larger batches directly impacts to the time which critical as it limits performance. While reducing the computational complexity in the trusted setup phase, such as by lowering the \(n\) value in the Power of \( \tau \) ceremony, could alleviate some of the pressure, we have to sacrifice the security then (trilemma). Moreover, exploring lighter ZKP protocols could offer more efficient solutions. Additionally, designing specialized hardware could help mitigate the computational bottlenecks observed during Power of Tau and key generation, further enhancing overall system performance \cite{daftardar2024szkp}.

Another potential direction for exploration is to analyze the real time ZKRollup solutions \cite{chaliasos2024analyzing}. However, it is important to note that executing the Polygon ZKEVM \cite{polygonzkEVM} solution requires approximately 1TB of main memory  \cite{chaliasos2024zkrollup}. It is also possible to benchmark other performance metrics such as memory usage, throughput, network load, bandwidth consumption, latency, etc. These metrics provide valuable insights into the system’s efficiency and scalability, helping to assess its overall performance under different conditions.

\section{Conclusion}
This paper presents an in-depth performance analysis of a ZKRollup solution using the Hardhat Ethereum development environment, focusing on both one-time and repetitive costs. We have demonstrated how larger batch sizes lead to increased setup costs due to the added complexity of generating cryptographic keys, compiling circuits, and running the Power of Tau ceremony. However, once the initial setup is completed, repetitive costs, such as proof generation and verification, scale more favorably, with the cost per transaction decreasing due to economies of scale. Our findings underline the potential of ZKRollup as a scalable solution for Ethereum, with room for optimization in proof generation process.

\bibliography{main}
\bibliographystyle{IEEEtran} 

\end{document}